\def\Fint{\rlap{$\Biggl\rfloor$}\Biggl\lceil}

\documentclass[12pt]{article}

\begin{document}
\begin{titlepage}
\begin{flushright}
UFIFT-QG-04-4 \\ gr-qc/0411003
\end{flushright}
\vspace{.4cm}
\begin{center}
\textbf{Stress Tensor Correlators in the Schwinger-Keldysh Formalism}
\end{center}
\begin{center}
L. H. Ford$^{\dagger}$
\end{center}
\begin{center}
\textit{Institute of Cosmology \\ Department of Physics and Astronomy \\ 
Tufts University \\ Medford, MA 02155 USA}
\end{center}

\begin{center}
R. P. Woodard$^{\ddagger}$
\end{center}
\begin{center}
\textit{Department of Physics \\ University of Florida \\
Gainesville, FL 32611 USA}
\end{center}

\begin{center}
ABSTRACT
\end{center}
We express stress tensor correlators using the Schwinger-Keldysh formalism.
The absence of off-diagonal counterterms in this formalism ensures that the
$+-$ and $-+$ correlators are free of primitive divergences. We use 
dimensional regularization in position space to explicitly check this at 
one loop order for a massless scalar on a flat space background. We use the 
same procedure to show that the $++$ correlator contains the divergences 
first computed by `t Hooft and Veltman for the scalar contribution to the 
graviton self-energy.

\begin{flushleft}
PACS numbers: 4.62.+v, 04.60.-m, 05.40.-a
\end{flushleft}
\vspace{.4cm}
\begin{flushleft}
$^{\dagger}$ e-mail: ford@cosmos.phy.tufts.edu \\
$^{\ddagger}$ e-mail: woodard@phys.ufl.edu
\end{flushleft}
\end{titlepage}

\section{Introduction}

Quantum fluctuations of the stress tensor operator play a role in 
at least three seemingly distinct physical phenomena: fluctuations of 
the Casimir force, radiation pressure fluctuations, 
and passive fluctuations of the gravitational field.
The Casimir force between a pair of material bodies is a mean force,
which can be computed from the expectation value of the stress tensor.
However, fluctuations around this mean value are expected, and have been
discussed by several authors~\cite{Barton,Eberlein,JR,WKF01}. 
Unfortunately, these fluctuations
seem to be too small to be observable at the present time.

Quantum fluctuations of the radiation pressure can also be
interpreted as a manifestation of quantum stress tensor 
fluctuations~\cite{WF01}.
Although this effect has not yet been observed, it is likely to be 
detected as part of the future development of laser interferometer
detectors of gravity waves.  

Just as the stress tensor describes forces on material bodies, it also
acts as the source of the gravitational field in general relativity.
The semiclassical theory assumes that this source is the expectation
value of the  stress tensor. This is an approximation which fails
when the stress tensor fluctuations are significant, which can occur
far from the Planck scale~\cite{F82, Kuo}. In this case, there are large
fluctuations of the gravitational field around the mean value predicted
by the semiclassical theory. Among the physical effects produced by
such fluctuations are the angular blurring and luminosity
fluctuations of a distance source~\cite{BF}. Other effects of stress
tensor fluctuations might play a role 
in the early universe, or near evaporating black 
holes~\cite{CH95,TW,CCV,MV99,HS98,BFP00}.   

These physical effects involve observables which are expressible as 
spacetime integrals of a stress tensor correlation function,
or correlator. This correlator (also known as the noise kernel) 
is a function of two spacetime points,
$x$ and $x'$, which has a $(x-x')^{-8}$ singularity as 
$x' \rightarrow x$ in four dimensions. As a result, the spacetime
integrals are formally divergent and must be regularized. One approach
which has been employed is an integration by parts procedure~\cite{WF01,FW03},
which is essentially differential regularization~\cite{FJL}. In many
cases, this procedure leads to a finite result without the need for any
renormalization. 

In this paper, we will examine the ultraviolet singularities of the
stress tensor correlators using dimensional regularization in position
space and the Schwinger-Keld\-ysh formalism. This will lead to insight 
as to when renormalization is required and when it is not. In 
Sec.~\ref{sec:SK}, we will review the  Schwinger-Keldysh formalism. In 
Sec.~\ref{sec:III}, the correlators for a massless scalar field will be 
constructed, and their ultraviolet singularities examined. Finally, in 
Sec.~\ref{sec:final}, we discuss the results.

\section{The Schwinger-Keldysh Formalism}
\label{sec:SK}

The Schwinger-Keldysh formalism is a technique that makes computing
expectation values almost as simple as the Feynman rules do for computing
in-out matrix elements \cite{JS,M,BM,K,CSHY,RJ,CH}. To sketch the 
derivation, consider a real scalar field, $\varphi(x)$ whose Lagrangian 
(not Lagrangian density) at time $t$ is $L[\varphi(t)]$. The well-known 
functional integral expression for the matrix element of an operator 
$\mathcal{O}_1[ \varphi]$ between states whose wave functionals are given 
at a starting time $s$ and a last time $\ell$ is
\begin{equation}
\Bigl\langle \Phi \Bigl\vert T^*\Bigl(\mathcal{O}_1[\varphi]\Bigr) \Bigr\vert
\Psi \Bigr\rangle = \Fint [d\varphi] \, \mathcal{O}_1[\varphi] \, 
\Phi^*[\varphi(\ell)] \, e^{i \int_{s}^{\ell} dt L[\varphi(t)]} \, 
\Psi[\varphi(s)] \; . \label{1stint}
\end{equation}
The $T^*$-ordering symbol in the matrix element indicates that the operator
$\mathcal{O}_1[\varphi]$ is time-ordered, except that any derivatives are 
taken {\it outside} the time-ordering. We can use (\ref{1stint}) to obtain a
similar expression for the matrix element of the {\it anti}-time-ordered 
product of some operator $\mathcal{O}_2[\varphi]$ in the presence of the 
reversed states,
\begin{eqnarray}
\Bigl\langle \Psi \Bigl\vert \overline{T}^*\Bigl(\mathcal{O}_2[\varphi]\Bigr) 
\Bigr\vert \Phi \Bigr\rangle & = & \Bigl\langle \Phi \Bigl\vert T^*\Bigl(
\mathcal{O}_2^{\dagger}[\varphi]\Bigr) \Bigr\vert \Psi \Bigr\rangle^* \; , \\
& = & \Fint [d\varphi] \, \mathcal{O}_2[\varphi] \, \Phi[\varphi(\ell)] \,
e^{-i \int_{s}^{\ell} dt L[\varphi(t)]} \, \Psi^*[\varphi(s)] \; . 
\label{2ndint}
\end{eqnarray}

Now note that summing over a complete set of states $\Phi$ gives a delta 
functional,
\begin{equation}
\sum_{\Phi} \Phi\Bigl[\varphi_-(\ell)\Bigr] \, \Phi^*\Bigl[\varphi_+(\ell)
\Bigr] = \delta\Bigl[\varphi_-(\ell) \!-\! \varphi_+(\ell) \Bigr] 
\; . \label{sum}
\end{equation}
Taking the product of (\ref{1stint}) and (\ref{2ndint}), and using (\ref{sum}),
we obtain a functional integral expression for the expectation value of any
anti-time-ordered operator $\mathcal{O}_2$ multiplied by any time-ordered 
operator $\mathcal{O}_1$,
\begin{eqnarray}
\lefteqn{\Bigl\langle \Psi \Bigl\vert \overline{T}^*\Bigl(\mathcal{O}_2[
\varphi]\Bigr) T^*\Bigl(\mathcal{O}_1[\varphi]\Bigr) \Bigr\vert \Psi 
\Bigr\rangle = \Fint [d\varphi_+] [d\varphi_-] \, \delta\Bigl[\varphi_-(\ell) 
\!-\! \varphi_+(\ell)\Bigr] } \nonumber \\
& & \hspace{1.5cm} \times \mathcal{O}_2[\varphi_-] \mathcal{O}_1[\varphi_+] 
\Psi^*[\varphi_-(s)] e^{i \int_s^{\ell} dt \Bigl\{L[\varphi_+(t)] - 
L[\varphi_-(t)]\Bigr\}} \Psi[\varphi_+(s)] \; . \qquad \label{fund}
\end{eqnarray}
This is the fundamental relation between the canonical operator formalism
and the functional integral formalism in the Schwinger-Keldysh formalism.

The Feynman rules follow from (\ref{fund}) in close analogy to those for
in-out matrix elements. Because the same field is represented by two different
dummy functional variables, $\varphi_{\pm}(x)$, the endpoints of lines carry
a $\pm$ polarity. External lines associated with the operator $\mathcal{O}_2[
\varphi]$ have $-$ polarity whereas those associated with the operator 
$\mathcal{O}_1[\varphi]$ have $+$ polarity. Interaction vertices are either
all $+$ or all $-$. Vertices with $+$ polarity are the same as in the usual 
Feynman rules whereas vertices with the $-$ polarity have an additional minus 
sign. Propagators can be $++$, $-+$, $+-$ and $--$.

The four propagators can be read off from the fundamental relation (\ref{fund}) 
when the free Lagrangian is substituted for the full one. It is useful to
denote canonical expectation values in the free theory with a subscript $0$. 
With this convention we see that the $++$ propagator is just the ordinary 
Feynman propagator,
\begin{equation}
i\Delta_{\scriptscriptstyle ++}(x;x') = \Bigl\langle \Omega \Bigl\vert
T\Bigl(\varphi(x) \varphi(x') \Bigr) \Bigr\vert \Omega \Bigr\rangle_0 =
i\Delta(x;x') \; . \label{++}
\end{equation}
The other cases are simple to read off and to relate to the Feynman propagator,
\begin{eqnarray}
i\Delta_{\scriptscriptstyle -+}(x;x') \!\!\! & = & \!\!\! \Bigl\langle \Omega 
\Bigl\vert \varphi(x) \varphi(x') \Bigr\vert \Omega \Bigr\rangle_0 = 
\theta(t\!-\!t') i\Delta(x;x') \!+\! \theta(t'\!-\!t) \Bigl[i\Delta(x;x')
\Bigr]^* \! , \quad \label{-+} \\
i\Delta_{\scriptscriptstyle +-}(x;x') \!\!\! & = & \!\!\! \Bigl\langle \Omega 
\Bigl\vert \varphi(x') \varphi(x) \Bigr\vert \Omega \Bigr\rangle_0 =
\theta(t\!-\!t') \Bigl[i\Delta(x;x')\Bigr]^* \!\!+\! \theta(t'\!-\!t) 
i\Delta(x;x') , \quad \label{+-} \\
i\Delta_{\scriptscriptstyle --}(x;x') \!\!\! & = & \!\!\! \Bigl\langle \Omega 
\Bigl\vert \overline{T}\Bigl(\varphi(x) \varphi(x') \Bigr) \Bigr\vert 
\Omega \Bigr\rangle_0 = \Bigl[i\Delta(x;x')\Bigr]^* . \label{--}
\end{eqnarray}
Therefore we can get the four propagators of the Schwinger-Keldysh formalism
from the Feynman propagator once that is known.

\section{Massless Scalar Stress Tensor Correlators}
\label{sec:III}

The Lagrangian density for a massless, minimally coupled scalar $\varphi$ 
in the presence of an arbitrary spacelike metric $g_{\mu\nu}$ is,
\begin{equation}
\mathcal{L} = -\frac12 \partial_{\mu} \varphi \partial_{\nu} \varphi
g^{\mu\nu} \sqrt{-g} \; . \label{Lag}
\end{equation}
Specializing its stress tensor to flat space $g_{\mu\nu} = \eta_{\mu\nu}$ 
gives,
\begin{equation}
T_{\mu\nu} \equiv -\frac2{\sqrt{-g}} \frac{\delta S[\varphi,g]}{\delta g^{
\mu\nu}}\Biggl\vert_{g = \eta} \!\!\! = \;\; \partial_{\mu} \varphi 
\partial_{\nu} \varphi - \frac12 \eta_{\mu\nu} \eta^{\rho \sigma} 
\partial_{\rho} \varphi \partial_{\sigma} \varphi \; .
\end{equation}
From the discussion of the preceding section we see that there are four
natural 2-point correlators of this operator in the Schwinger-Keldysh
formalism,
\begin{eqnarray}
\Bigl[{}^{~+}_{\mu\nu} \mathcal{C}^+_{\rho\sigma}\Bigr](x;x') & \equiv & 
\Bigl\langle \Omega \Bigl\vert T\Bigl( T_{\mu\nu}(x) T_{\rho \sigma}(x') 
\Bigr) \Bigr\vert \Omega \Bigr\rangle \; , \label{C++} \\
\Bigl[{}^{~-}_{\mu\nu} \mathcal{C}^+_{\rho\sigma}\Bigr](x;x') & \equiv & 
\Bigl\langle \Omega \Bigl\vert \overline{T}\Bigl( T_{\mu\nu}(x)\Bigr) 
T\Bigl( T_{\rho \sigma}(x') \Bigr) \Bigr\vert \Omega \Bigr\rangle \label{C-+}\\ 
&=& \Bigl\langle \Omega \Bigl\vert  T_{\mu\nu}(x) 
 T_{\rho \sigma}(x') \Bigr\vert \Omega \Bigr\rangle  \; , \nonumber \\
\Bigl[{}^{~+}_{\mu\nu} \mathcal{C}^-_{\rho\sigma}\Bigr](x;x') & \equiv & 
\Bigl\langle \Omega \Bigl\vert \overline{T}\Bigl( T_{\rho\sigma}(x')\Bigr) 
T\Bigl( T_{\mu \nu}(x) \Bigr) \Bigr\vert \Omega \Bigr\rangle \label{C+-} \\ 
&=&  \Bigl\langle \Omega \Bigl\vert T_{\rho \sigma}(x') T_{\mu\nu}(x) 
  \Bigr\vert \Omega \Bigr\rangle \; , \nonumber \\
\Bigl[{}^{~-}_{\mu\nu} \mathcal{C}^-_{\rho\sigma}\Bigr](x;x') & \equiv & 
\Bigl\langle \Omega \Bigl\vert \overline{T}\Bigl( T_{\mu\nu}(x) T_{\rho 
\sigma}(x') \Bigr) \Bigr\vert \Omega \Bigr\rangle \; . \label{C--}
\end{eqnarray}
These four quantities are closely related. For example, note that (\ref{C++})
and (\ref{C--}) are complex conjugates, as are (\ref{C-+}) and (\ref{C+-}).
They have slightly different uses and physical interpretations. For example,
$-i 4 \pi G$ times (\ref{C++}) gives the scalar contribution to the graviton 
self-energy whose divergent part was computed by `t Hooft and Veltman 
\cite{HV}. The stress tensor fluctuations whose effect upon focusing has been 
studied recently \cite{BF} are given by $+\frac12$ times the sum of 
(\ref{C-+}) and (\ref{C+-}).

Each of the four Schwinger-Keldysh scalar propagators takes the same form in
$D$-dimensional flat space,
\begin{equation}
i \Delta_{\scriptscriptstyle \pm \pm}(x;x') = \frac{\Gamma(\frac{D}2 \!-\!1)}{
4 \pi^{\frac{D}2}} \left(\frac1{\Delta x_{\scriptscriptstyle \pm \pm}^2}
\right)^{\frac{D}2 - 1} \; .
\end{equation}
The four $\pm$ variations only affect what we mean by the invariant interval,
\begin{eqnarray}
\Delta x^2_{\scriptscriptstyle ++} \equiv \Bigl\Vert \vec{x} \!-\! \vec{x}' 
 \Bigr\Vert^2 - \Bigl(\vert t \!-\! t'\vert \!-\! i \delta \Bigr)^2 & , &
\Delta x^2_{\scriptscriptstyle +-} \equiv \Bigl\Vert \vec{x} \!-\! \vec{x}' 
\Bigr\Vert^2 - \Bigl(t \!-\! t' \!+\! i \delta \Bigr)^2 , \nonumber \\
\Delta x^2_{\scriptscriptstyle --} \equiv \Bigl\Vert \vec{x} \!-\! \vec{x}' 
\Bigr\Vert^2 - \Bigl(\vert t \!-\! t'\vert \!+\! i \delta \Bigr)^2 & , &
\Delta x^2_{\scriptscriptstyle -+} \equiv \Bigl\Vert \vec{x} \!-\! \vec{x}' 
\Bigr\Vert^2 - \Bigl(t \!-\! t' \!-\! i \delta \Bigr)^2 . \qquad
\end{eqnarray}
Because the $+-$ and $-+$ intervals involve $t\!-\!t'$, rather than $\vert
t \!-\! t'\vert$, second derivatives of these propagators are straightforward,
\begin{eqnarray}
\partial_{\kappa} \partial_{\alpha}' i\Delta_{\scriptscriptstyle +-}(x;x') &=& 
\frac{\Gamma(\frac{D}2)}{2 \pi^{\frac{D}2}} \left[ \frac{\eta_{\kappa\alpha}}{
\Delta x^D_{\scriptscriptstyle +-}} - D \frac{\Delta x_{\kappa} \Delta x_{
\alpha}}{\Delta x^{D+2}_{\scriptscriptstyle +-}} \right] \; , \\
\partial_{\kappa} \partial_{\alpha}' i\Delta_{\scriptscriptstyle -+}(x;x') &=& 
\frac{\Gamma(\frac{D}2)}{2 \pi^{\frac{D}2}} \left[ \frac{\eta_{\kappa\alpha}}{
\Delta x^D_{\scriptscriptstyle -+}} - D \frac{\Delta x_{\kappa} \Delta x_{
\alpha}}{\Delta x^{D+2}_{\scriptscriptstyle -+}} \right] \; .
\end{eqnarray}
Second derivatives of the $++$ and $--$ propagators involve another term 
owing to the absolute value \cite{OW,PTW},
\begin{eqnarray}
\partial_{\kappa} \partial_{\alpha}' i\Delta_{\scriptscriptstyle ++}(x;x') &=&
\frac{\Gamma(\frac{D}2)}{2 \pi^{\frac{D}2}} \left[ \frac{\eta_{\kappa\alpha}}{
\Delta x^D_{\scriptscriptstyle ++}} - D \frac{\Delta x_{\kappa} \Delta x_{
\alpha}}{\Delta x^{D+2}_{\scriptscriptstyle ++}} \right] + i \delta^0_{\kappa}
\delta^0_{\alpha} \delta^D(x \!-\! x') \; , \\
\partial_{\kappa} \partial_{\alpha}' i\Delta_{\scriptscriptstyle --}(x;x') &=&
\frac{\Gamma(\frac{D}2)}{2 \pi^{\frac{D}2}} \left[ \frac{\eta_{\kappa\alpha}}{
\Delta x^D_{\scriptscriptstyle --}} - D \frac{\Delta x_{\kappa} \Delta x_{
\alpha}}{\Delta x^{D+2}_{\scriptscriptstyle --}} \right] - i \delta^0_{\kappa}
\delta^0_{\alpha} \delta^D(x \!-\! x') \; .
\end{eqnarray}
However, note that this extra term goes away when the derivatives act inside
the time-ordering (or anti-time-ordering) symbol,
\begin{equation}
\Bigl\langle \Omega \Bigl\vert T\Bigl(\partial_{\kappa} \varphi(x) 
\partial_{\alpha}' \varphi(x') \Bigr) \Bigr\vert \Omega \Bigr\rangle =
\frac{\Gamma(\frac{D}2)}{2 \pi^{\frac{D}2}} \left[ \frac{\eta_{\kappa\alpha}}{
\Delta x^D_{\scriptscriptstyle ++}} - D \frac{\Delta x_{\kappa} \Delta x_{
\alpha}}{\Delta x^{D+2}_{\scriptscriptstyle ++}} \right] \; .
\end{equation}
Note also that the coincidence limits of such quantities vanish in dimensional
regularization \cite{OW,PTW},
\begin{equation}
\Bigl\langle \Omega \Bigl\vert T\Bigl(\partial_{\kappa} \varphi(x) 
\partial_{\lambda} \varphi(x) \Bigr) \Bigr\vert \Omega \Bigr\rangle = 0 =
\Bigl\langle \Omega \Bigl\vert \overline{T}\Bigl(\partial_{\kappa} \varphi(x) 
\partial_{\lambda} \varphi(x) \Bigr) \Bigr\vert \Omega \Bigr\rangle \; .
\end{equation}

We can now evaluate the various correlators quite simply. Consider first 
the $++$ case,
\begin{eqnarray}
\lefteqn{\Bigl\langle \Omega \Bigl\vert T\Bigl( T_{\mu\nu}(x) T_{\rho 
\sigma}(x') \Bigr) \Bigr\vert \Omega \Bigr\rangle } \nonumber \\
& & = \Bigl[ \delta^{\kappa}_{\mu} \delta^{\lambda}_{\nu} \!-\! \frac12 
\eta_{\mu\nu} \eta^{\kappa\lambda} \Bigr] \Bigl[ \delta^{\alpha}_{\rho} 
\delta^{\beta}_{\sigma} \!-\! \frac12 \eta_{\rho\sigma} \eta^{\alpha\beta} 
\Bigr] \Bigl\langle \Omega \Bigl\vert T\Bigl(\partial_{\kappa} \varphi 
\partial_{\lambda} \varphi \partial_{\alpha}' \varphi \partial_{\beta}' 
\varphi \Bigr) \Bigr\vert \Omega \Bigr\rangle , \\
& & = \Bigl[ \delta^{\kappa}_{(\mu} \delta^{\lambda}_{\nu)} \!-\! \frac12 
\eta_{\mu\nu} \eta^{\kappa\lambda} \Bigr] \Bigl[ \delta^{\alpha}_{(\rho} 
\delta^{\beta}_{\sigma)} \!-\! \frac12 \eta_{\rho\sigma} \eta^{\alpha\beta} 
\Bigr] \nonumber \\
& & \hspace{4cm} \times 2 \Bigl\langle \Omega \Bigl\vert T\Bigl(\partial_{
\kappa} \varphi \partial_{\alpha}' \varphi \Bigl) \Bigr\vert \Omega 
\Bigr\rangle \Bigl\langle \Omega \Bigl\vert T\Bigl(\partial_{\lambda} \varphi 
\partial_{\beta}' \varphi \Bigl) \Bigr\vert \Omega \Bigr\rangle , \\
& & = \frac{\Gamma^2(\frac{D}2)}{2 \pi^D} \left\{ \frac{\eta_{\mu ( \rho}
\eta_{\sigma) \nu}}{\Delta x_{\scriptscriptstyle ++}^{2D}} - 2D \frac{
\Delta x_{(\mu} \eta_{\nu) (\rho} \Delta x_{\sigma)}}{\Delta x_{
\scriptscriptstyle ++}^{2D+2}} + D^2 \frac{\Delta x_{\mu} \Delta x_{\nu}
\Delta x_{\rho} \Delta x_{\sigma}}{\Delta x_{\scriptscriptstyle ++}^{2D+4}}
\right. \nonumber \\
& & \hspace{.2cm} \left. -\frac12 (D\!-\!2) D \Bigl[\frac{\eta_{\mu \nu}
\Delta x_{\rho} \Delta x_{\sigma} \!+\! \eta_{\rho\sigma} \Delta x_{\mu} 
\Delta x_{\nu}}{\Delta x_{\scriptscriptstyle ++}^{2D+2}} \Bigr] \!+\! \frac14
(D^2 \!-\! D \!-\! 4) \frac{\eta_{\mu\nu} \eta_{\rho\sigma}}{\Delta x_{
\scriptscriptstyle ++}^{2D}} \right\} . \qquad
\end{eqnarray}
Although the intermediate steps are different, the result takes the
same form for all four correlators,
\begin{eqnarray}
\lefteqn{ \Bigl[{}^{~\pm}_{\mu\nu} \mathcal{C}^\pm_{\rho\sigma}\Bigr](x;x') } 
\nonumber \\
& & = \frac{\Gamma^2(\frac{D}2)}{2 \pi^D} \left\{ \frac{\eta_{\mu ( \rho}
\eta_{\sigma) \nu}}{\Delta x_{\scriptscriptstyle \pm\pm}^{2D}} - 2D \frac{
\Delta x_{(\mu} \eta_{\nu) (\rho} \Delta x_{\sigma)}}{\Delta x_{
\scriptscriptstyle \pm\pm}^{2D+2}} + D^2 \frac{\Delta x_{\mu} \Delta x_{\nu}
\Delta x_{\rho} \Delta x_{\sigma}}{\Delta x_{\scriptscriptstyle \pm\pm}^{2D+4}}
\right. \nonumber \\
& & \hspace{.2cm} \left. -\frac12 (D\!-\!2) D \Bigl[\frac{\eta_{\mu \nu}
\Delta x_{\rho} \Delta x_{\sigma} \!+\! \eta_{\rho\sigma} \Delta x_{\mu} 
\Delta x_{\nu}}{\Delta x_{\scriptscriptstyle \pm\pm}^{2D+2}} \Bigr] \!+\! 
\frac14 (D^2 \!-\! D \!-\! 4) \frac{\eta_{\mu\nu} \eta_{\rho\sigma}}{\Delta x_{
\scriptscriptstyle \pm\pm}^{2D}} \right\} . \qquad
\end{eqnarray}

The next step is to partially integrate up to the logarithmically divergent
power $1/{\Delta x}^{2D-4}$. This is facilitated by the following 
identities \cite{OW,PTW,PW},
\begin{eqnarray}
\frac1{\Delta x_{\scriptscriptstyle \pm\pm}^{2D}} &=& \left\{\frac{\partial^4}{
4 (D \!-\!2)^2 (D \!-\! 1)D} \right\} \frac1{\Delta x_{\scriptscriptstyle 
\pm\pm}^{2D-4}} , \label{D1} \\
\frac{\Delta x_{\mu} \Delta x_{\nu}}{\Delta x_{\scriptscriptstyle 
\pm\pm}^{2D+2}} &=& \left\{ \frac{\eta_{\mu\nu} \partial^4 \!+\! D \partial_{
\mu} \partial_{\nu} \partial^2}{8 (D \!-\!2)^2 (D \!-\! 1) D^2} \right\} 
\frac1{\Delta x_{\scriptscriptstyle \pm\pm}^{2D-4}} , \label{D2} \\
\frac{\Delta x_{\mu} \Delta x_{\nu} \Delta x_{\rho} \Delta x_{\sigma}}{
\Delta x_{\scriptscriptstyle \pm\pm}^{2D+4}} &=& \left\{ \frac{\Bigl[\eta_{\mu
\nu} \eta_{\rho\sigma} \!+\! 2 \eta_{\mu (\rho} \eta_{\sigma) \nu}\Bigr]
\partial^4}{16 (D \!-\!2)^2 (D \!-\! 1)D^2 (D \!+\! 1)} \right. \nonumber \\
& & \hspace{-3.5cm} \left. + \frac{\Bigl[\eta_{\mu \nu} \partial_{\rho} 
\partial_{\sigma} \!+\! 4 \partial_{(\mu} \eta_{\nu) (\rho} \partial_{\sigma)} 
\!+\! \eta_{\rho\sigma} \partial_{\mu} \partial_{\nu}\Bigr] \partial^2 \!+\! 
(D \!-\! 2) \partial_{\mu} \partial_{\nu} \partial_{\rho} \partial_{\sigma}}{
16 (D \!-\!2)^2 (D \!-\! 1) D (D \!+\! 1)} \right\} \frac1{\Delta x_{
\scriptscriptstyle \pm\pm}^{2D-4}} . \qquad \label{D3}
\end{eqnarray}
At this stage the result takes the form,
\begin{eqnarray}
\lefteqn{ \Bigl[{}^{~\pm}_{\mu\nu} \mathcal{C}^\pm_{\rho\sigma}\Bigr](x;x') =
\frac{\Gamma^2(\frac{D}2)}{16 \pi^D} \left\{
\frac{\Bigl(D^2 \!-\! 2D \!-\! 2\Bigr) \Bigl[\eta_{\mu\nu} \partial^2 \!-\! 
\partial_{\mu} \partial_{\nu}\Bigr] \Bigl[\eta_{\rho\sigma} \partial^2 \!-\! 
\partial_{\rho} \partial_{\sigma}\Bigr]}{2 (D\!-\!2)^2 (D\!-\!1) (D\!+\!1)} 
\right. } \nonumber \\
& & \hspace{2.5cm} \left. + \frac{\Bigl[\eta_{\mu (\rho} \eta_{\sigma) \nu} 
\partial^4 \!-\! 2 \partial_{(\mu} \eta_{\nu) (\rho} \partial_{\sigma)} 
\partial^2 \!+\!  \partial_{\mu} \partial_{\nu} \partial_{\rho} 
\partial_{\sigma}\Bigr]}{(D\!-\!2)^2 (D\!-\!1) (D\!+\!1)} \right\} 
\frac1{\Delta x_{\scriptscriptstyle \pm\pm}^{2D-4}} . \qquad \label{step2}
\end{eqnarray}
Note the manifest transversality of (\ref{step2}) which is a consequence
of stress-energy conservation.

At this point we pause to note that no delta functions emerge from the 
derivatives in (\ref{D1}-\ref{D3}) because the only power that can give 
them in dimensional regularization is $1/{\Delta x}^{D-2}$. This happens 
for the $++$ and $--$ cases \cite{OW,PTW,PW},
\begin{equation}
\partial^2 \left( \frac1{\Delta x_{\scriptscriptstyle ++}^{D-2}} \right)
= \frac{i 4 \pi^{\frac{D}2}}{\Gamma(\frac{D}2 \!-\! 1)} \, \delta^D(x \!-\! x')
= - \partial^2 \left( \frac1{\Delta x_{\scriptscriptstyle --}^{D-2}} \right) .
\label{delt}
\end{equation}
It does not happen for the $+-$ and $-+$ cases \cite{OW,PTW,PW},
\begin{equation}
\partial^2 \left( \frac1{\Delta x_{\scriptscriptstyle +-}^{D-2}} \right) = 0 
= \partial^2 \left( \frac1{\Delta x_{\scriptscriptstyle -+}^{D-2}} \right) .
\label{nodelt}
\end{equation}

The point of partially integrating, as we did to reach (\ref{step2}), is
to write the result as a derivative operator with respect to $x^{\mu}$,
acting upon a function of $x^{\prime \mu}$ which is integrable in $D=4$. We
have not quite achieved this in (\ref{step2}) but the next partial
integration does,
\begin{equation}
\frac1{\Delta x_{\scriptscriptstyle \pm\pm}^{2D-4}} = \frac{\partial^2}{2 (D 
\!-\! 3) (D \!-\! 4)} \left( \frac1{\Delta x_{\scriptscriptstyle \pm\pm}^{
2D-6}} \right) \; .
\end{equation}
Except for the explicit factor of $1/(D\!-\!4)$ we could take $D=4$ in this
expression. 

The next step --- and the first at which we must distinguish between the 
four $\pm$ variations --- is to transfer the divergence to a local term by 
adding zero in the form of the identities (\ref{delt}) and (\ref{nodelt}). 
For the $++$ term this gives,
\begin{eqnarray}
\lefteqn{\frac1{\Delta x_{\scriptscriptstyle ++}^{2D-4}} = \frac{\partial^2}{2 
(D \!-\! 3) (D \!-\! 4)} \left( \frac1{\Delta x_{\scriptscriptstyle ++}^{2D-6}} 
\!-\! \frac{\mu^{D-4}}{\Delta x_{\scriptscriptstyle ++}^{D-2}} \right) } 
\nonumber \\
& & \hspace{5cm} + \frac{i 4 \pi^{\frac{D}2} \mu^{D-4}}{\Gamma(\frac{D}2 \!-\! 
1)} \, \frac{\delta^D(x \!-\! x')}{2 (D \!-\! 3) (D \!-\! 4)} \; . \label{loc}
\end{eqnarray}
Note the dimensional regularization mass scale $\mu$. The expression on 
the first line of (\ref{loc}) is both integrable {\it and} finite so we
can take $D=4$,
\begin{equation}
\frac{\partial^2}{2 (D \!-\! 3) (D \!-\! 4)} \left( \frac1{\Delta x_{
\scriptscriptstyle ++}^{2D-6}} \!-\! \frac{\mu^{D-4}}{\Delta x_{
\scriptscriptstyle ++}^{D-2}} \right) \longrightarrow -\frac{\partial^2}4
\left(\frac{\ln(\mu^2 \Delta x_{\scriptscriptstyle ++}^2)}{ \Delta 
x_{\scriptscriptstyle ++}^2} \right) \; .
\end{equation}
The result for each of the four $\pm$ variations is,
\begin{eqnarray}
\frac1{\Delta x_{\scriptscriptstyle ++}^{2D-4}} & \longrightarrow & 
-\frac{\partial^2}4 \left(\frac{\ln(\mu^2 \Delta x_{\scriptscriptstyle ++}^2)}{
\Delta x_{\scriptscriptstyle ++}^2} \right) 
+ \frac{i 4 \pi^{\frac{D}2} \mu^{D-4}}{\Gamma(\frac{D}2 \!-\! 
1)} \, \frac{\delta^D(x \!-\! x')}{2 (D \!-\! 3) (D \!-\! 4)} \; , \\
\frac1{\Delta x_{\scriptscriptstyle +-}^{2D-4}} & \longrightarrow & 
-\frac{\partial^2}4 \left(\frac{\ln(\mu^2 \Delta x_{\scriptscriptstyle +-}^2)}{
\Delta x_{\scriptscriptstyle +-}^2} \right) \; , \\
\frac1{\Delta x_{\scriptscriptstyle -+}^{2D-4}} & \longrightarrow & 
-\frac{\partial^2}4 \left(\frac{\ln(\mu^2 \Delta x_{\scriptscriptstyle -+}^2)}{
\Delta x_{\scriptscriptstyle -+}^2} \right) \; , \\
\frac1{\Delta x_{\scriptscriptstyle --}^{2D-4}} & \longrightarrow & 
-\frac{\partial^2}4 \left(\frac{\ln(\mu^2 \Delta x_{\scriptscriptstyle --}^2)}{
\Delta x_{\scriptscriptstyle --}^2} \right) 
- \frac{i 4 \pi^{\frac{D}2} \mu^{D-4}}{\Gamma(\frac{D}2 \!-\! 
1)} \, \frac{\delta^D(x \!-\! x')}{2 (D \!-\! 3) (D \!-\! 4)} \; .
\end{eqnarray}

We see that the $+-$ and $-+$ correlators are completely finite,
\begin{eqnarray}
\lefteqn{\Bigl[{}^{~+}_{\mu\nu} \mathcal{C}^-_{\rho\sigma}\Bigr](x;x') 
\longrightarrow \frac{-\partial^2}{1280 \pi^4} \Biggl\{ \Bigl[\eta_{\mu\nu} 
\partial^2 \!-\! \partial_{\mu} \partial_{\nu}\Bigr] \Bigl[\eta_{\rho\sigma} 
\partial^2 \!-\! \partial_{\rho} \partial_{\sigma}\Bigr] } \nonumber \\
& & \hspace{1.5cm} + \frac13 \Bigl[\eta_{\mu (\rho} \eta_{\sigma) \nu} 
\partial^4 \!-\! 2 \partial_{(\mu} \eta_{\nu) (\rho} \partial_{\sigma)} 
\partial^2 \!+\!  \partial_{\mu} \partial_{\nu} \partial_{\rho} \partial_{
\sigma}\Bigr]\Biggr\} \frac{\ln(\mu^2 \Delta x_{\scriptscriptstyle +-}^2)}{
\Delta x_{\scriptscriptstyle +-}^2} , \qquad \label{CF+-} \\
\lefteqn{\Bigl[{}^{~-}_{\mu\nu} \mathcal{C}^+_{\rho\sigma}\Bigr](x;x') 
\longrightarrow \frac{-\partial^2}{1280 \pi^4} \Biggl\{ \Bigl[\eta_{\mu\nu} 
\partial^2 \!-\! \partial_{\mu} \partial_{\nu}\Bigr] \Bigl[\eta_{\rho\sigma} 
\partial^2 \!-\! \partial_{\rho} \partial_{\sigma}\Bigr] } \nonumber \\
& & \hspace{1.5cm} + \frac13 \Bigl[\eta_{\mu (\rho} \eta_{\sigma) \nu} 
\partial^4 \!-\! 2 \partial_{(\mu} \eta_{\nu) (\rho} \partial_{\sigma)} 
\partial^2 \!+\!  \partial_{\mu} \partial_{\nu} \partial_{\rho} \partial_{
\sigma}\Bigr] \Biggr\} \frac{\ln(\mu^2 \Delta x_{\scriptscriptstyle -+}^2)}{
\Delta x_{\scriptscriptstyle -+}^2} . \qquad \label{CF-+}
\end{eqnarray}
Since they are complex conjugates, their average is also real. The $++$ and
$--$ terms have similar finite parts but they also harbor ultraviolet
divergences,
\begin{eqnarray}
\lefteqn{\Bigl[{}^{~+}_{\mu\nu} \mathcal{C}^+_{\rho\sigma}\Bigr](x;x') 
\longrightarrow \frac{-\partial^2}{1280 \pi^4} \Biggl\{ \Bigl[\eta_{\mu\nu} 
\partial^2 \!-\! \partial_{\mu} \partial_{\nu}\Bigr] \Bigl[\eta_{\rho\sigma} 
\partial^2 \!-\! \partial_{\rho} \partial_{\sigma}\Bigr] } \nonumber \\
& & \hspace{1.5cm} + \frac13 \Bigl[\eta_{\mu (\rho} \eta_{\sigma) \nu} 
\partial^4 \!-\! 2 \partial_{(\mu} \eta_{\nu) (\rho} \partial_{\sigma)} 
\partial^2 \!+\!  \partial_{\mu} \partial_{\nu} \partial_{\rho} \partial_{
\sigma}\Bigr]\Biggr\} \frac{\ln(\mu^2 \Delta x_{\scriptscriptstyle ++}^2)}{
\Delta x_{\scriptscriptstyle ++}^2} \nonumber \\
& & + \frac{i\Gamma(\frac{D}2) \mu^{D-4}}{16 \pi^{\frac{D}2}} \left\{
\frac{\Bigl(D^2 \!-\! 2D \!-\! 2\Bigr) \Bigl[\eta_{\mu\nu} \partial^2 \!-\! 
\partial_{\mu} \partial_{\nu}\Bigr] \Bigl[\eta_{\rho\sigma} \partial^2 \!-\! 
\partial_{\rho} \partial_{\sigma}\Bigr]}{2 (D \!-\!4) (D \!-\! 3) (D\!-\!2) 
(D\!-\!1) (D\!+\!1)} \right. \nonumber \\
& & \hspace{2.5cm} \left. + \frac{\Bigl[\eta_{\mu (\rho} \eta_{\sigma) \nu} 
\partial^4 \!-\! 2 \partial_{(\mu} \eta_{\nu) (\rho} \partial_{\sigma)} 
\partial^2 \!+\!  \partial_{\mu} \partial_{\nu} \partial_{\rho} 
\partial_{\sigma}\Bigr]}{(D \!-\! 4) (D \!-\! 3) (D\!-\!2) (D\!-\!1) 
(D\!+\!1)} \right\} \delta^D(x \!-\! x') , \label{CF++} \qquad \\
\lefteqn{\Bigl[{}^{~-}_{\mu\nu} \mathcal{C}^-_{\rho\sigma}\Bigr](x;x') 
\longrightarrow \frac{-\partial^2}{1280 \pi^4} \Biggl\{ \Bigl[\eta_{\mu\nu} 
\partial^2 \!-\! \partial_{\mu} \partial_{\nu}\Bigr] \Bigl[\eta_{\rho\sigma} 
\partial^2 \!-\! \partial_{\rho} \partial_{\sigma}\Bigr] } \nonumber \\
& & \hspace{1.5cm} + \frac13 \Bigl[\eta_{\mu (\rho} \eta_{\sigma) \nu} 
\partial^4 \!-\! 2 \partial_{(\mu} \eta_{\nu) (\rho} \partial_{\sigma)} 
\partial^2 \!+\!  \partial_{\mu} \partial_{\nu} \partial_{\rho} \partial_{
\sigma}\Bigr] \Biggr\} \frac{\ln(\mu^2 \Delta x_{\scriptscriptstyle --}^2)}{
\Delta x_{\scriptscriptstyle --}^2} \nonumber \\
& & - \frac{i\Gamma(\frac{D}2) \mu^{D-4}}{16 \pi^{\frac{D}2}} \left\{
\frac{\Bigl(D^2 \!-\! 2D \!-\! 2\Bigr) \Bigl[\eta_{\mu\nu} \partial^2 \!-\! 
\partial_{\mu} \partial_{\nu}\Bigr] \Bigl[\eta_{\rho\sigma} \partial^2 \!-\! 
\partial_{\rho} \partial_{\sigma}\Bigr]}{2 (D \!-\!4) (D \!-\! 3) (D\!-\!2) 
(D\!-\!1) (D\!+\!1)} \right. \nonumber \\
& & \hspace{2.5cm} \left. + \frac{\Bigl[\eta_{\mu (\rho} \eta_{\sigma) \nu} 
\partial^4 \!-\! 2 \partial_{(\mu} \eta_{\nu) (\rho} \partial_{\sigma)} 
\partial^2 \!+\!  \partial_{\mu} \partial_{\nu} \partial_{\rho} 
\partial_{\sigma}\Bigr]}{(D \!-\! 4) (D \!-\! 3) (D\!-\!2) (D\!-\!1) 
(D\!+\!1)} \right\} \delta^D(x \!-\! x') , \qquad \label{CF--}
\end{eqnarray}
These four correlators have been previously evaluated at one loop order,
also using dimensional regularization but in momentum space, by Campos
and Verdaguer \cite{CV} and by Martin and Verdaguer \cite{MV}. 

\section{Discussion}
\label{sec:final}

We have expressed the stress tensor correlators of previous studies
\cite{BF} using the Schwinger-Keldysh formalism \cite{JS,M,BM,K,CSHY,RJ,CH}. 
In this language it is the average of the $+-$ and $-+$ correlators
(\ref{CF+-}-\ref{CF-+}) which has been the object of earlier study. 
However, the four $\pm$ variations are so closely related that a unified 
treatment was simple. The reduction procedure is by now familiar from 
analogous computations in a locally de Sitter background \cite{OW,PTW,PW,BOW}.

It might seem curious that the $+-$ and $-+$ correlators 
(\ref{CF+-}-\ref{CF-+}) are ultraviolet finite. That this must be so 
derives from the relation between stress tensor correlators and the 
graviton self-energy. To see this relation, define the graviton field 
$h_{\mu\nu}(x)$ by perturbing the full metric about flat space,
\begin{equation}
g_{\mu\nu}(x) \equiv \eta_{\mu\nu} + \kappa h_{\mu\nu}(x) \qquad , \qquad
{\rm where} \quad \kappa^2 \equiv 16 \pi G \; .
\end{equation}
As usual in perturbative quantum gravity we raise and lower indices with the
background metric, $\eta_{\mu\nu}$. The 3-point interaction between gravitons
and scalars can be read off from ({\ref{Lag}),
\begin{equation}
g^{\mu\nu} \sqrt{-g} = \eta^{\mu\nu} - \kappa \Bigl(h^{\mu\nu} - \frac12
\eta^{\mu\nu} h \Bigr) + O(\kappa^2) \quad \Longrightarrow \quad 
\mathcal{L}^{(3)} = \frac{\kappa}2 h^{\mu\nu} T_{\mu\nu} . \qquad
\end{equation}
Although there is a 4-point interaction, it makes no contribution to the
graviton self-energy because the coincident massless scalar propagator 
vanishes in dimensional regularization. We can therefore write the four
Schwinger-Keldysh self-energies in terms of the four correlators,
\begin{equation}
-i \Bigl[{}^{~\pm}_{\mu\nu} \Sigma^{\pm}_{\rho\sigma}\Bigr](x;x') =
\Bigl(\frac{\pm i \kappa}2\Bigr) \Bigl(\frac{\pm i \kappa}2\Bigr) 
\Bigl[{}^{~\pm}_{\mu\nu} \mathcal{C}^{\pm}_{\rho\sigma}\Bigr](x;x') \; .
\end{equation}

The reason that the $+-$ and $-+$ correlators are finite becomes clear when
we consider how the various correlators enter the Schwinger-Keldysh effective
action \cite{RJ,CV,PTW,PW},
\begin{eqnarray}
\lefteqn{\Gamma[g_+,g_-] = S[g_+] - S[g_-] + \frac{i \kappa^2}8 \int d^4x 
\int d^4x' } \nonumber \\
& & \hspace{-.5cm} \left\{\!\!
\matrix{ h_+^{\mu\nu}(x) \Bigl[{}^{~+}_{\mu\nu} \mathcal{C}^{+}_{\rho\sigma}
\Bigr](x;x') h_+^{\rho\sigma}(x') \!-\! h_+^{\mu\nu}(x) \Bigl[{}^{~+}_{\mu\nu} 
\mathcal{C}^{-}_{\rho\sigma}\Bigr](x;x') h_-^{\rho\sigma}(x') \cr
-h_-^{\mu\nu}(x) \Bigl[{}^{~-}_{\mu\nu} \mathcal{C}^{+}_{\rho\sigma}
\Bigr](x;x') h_+^{\rho\sigma}(x') \!+\! h_-^{\mu\nu}(x) \Bigl[{}^{~-}_{\mu\nu} 
\mathcal{C}^{-}_{\rho\sigma}\Bigr](x;x') h_-^{\rho\sigma}(x')} \!\! \right\}
\!+\! O(\kappa^3) . \quad
\end{eqnarray}
Counterterms derive from the ``classical'' actions, $S[g_+]$ and $-S[g_-]$, 
hence they can involve only all $+$ or all $-$ fields. It follows that the
mixed correlators can not harbor primitive divergences at any order. At one 
loop order the only divergences are primitive, so these correlators are simply 
finite.

The $++$ and $--$ correlators do harbor divergences. It is illuminating
express the $++$ pole term in invariant form,
\begin{eqnarray}
\lefteqn{\Gamma^{\scriptscriptstyle ++}_{\infty}[g] = \frac{-\kappa^2}{1280 
\pi^2} \frac1{D\!-\!4} \int d^4x \Biggl\{ \Bigl[h^{\mu~ ,\nu}_{~\mu ~~ \nu} 
\!-\!  h^{\mu\nu}_{~~ ,\mu\nu} \Bigr]^2 } \nonumber \\
& & \hspace{2cm} + \frac13 \Bigl[ h^{\rho\sigma , \mu}_{~~~~ \mu} h^{~~ , 
\nu}_{\rho \sigma ~~ \nu}  \!-\! 2 h^{\rho\sigma , \mu}_{~~~~ \sigma} 
h^{\nu}_{~ \rho , \mu\nu} \!+\! h^{\rho\sigma}_{~~ ,\rho\sigma} 
h^{\mu\nu}_{~~ ,\mu\nu}\Bigr] \Biggr\} + O(\kappa^3) , \qquad \\
& & = \frac{-1}{960 \pi^2} \frac1{D \!-\! 4} \int d^4x \sqrt{-g} \Biggl\{
\frac12 R^2 + R^{\mu\nu} R_{\mu\nu} \Biggr\} .
\end{eqnarray}
This is exactly $-\frac12$ of the counterterm --- their equation (3.34) ---
that `t Hooft and Veltman long ago computed for removing the one loop 
divergences induced by a complex scalar \cite{HV}. Because a complex scalar 
contributes as two real scalars, our result is in perfect agreement, as it
is with subsequent studies \cite{CV,MV}.

We can also now understand why the integration by parts procedure used
Refs.~\cite{WF01,BF,FW03} leads to a finite result with no infinite
subtraction required. These papers were concerned solely with the
$+-$ and $-+$ correlators, which we have shown to be finite in
dimensional regularization. Even if dimensional regularization is
not used explicitly, integration by parts will produce a finite result.
   
Although the focus of this paper has been understanding the singularity 
structure of stress tensor correlators, the motivation for the exercise
is the interesting physics associated with these objects. In flat space
the physical effects arising from quantum stress tensor fluctuations 
are associated with the finite, state dependent parts of the correlator. 
Examples include radiation pressure fluctuations for an electromagnetic 
field in a coherent state \cite{WF01} and the angular blurring and 
luminosity fluctuations of the image of a distant source produced by 
Ricci tensor fluctuations, which can in turn arise from stress tensor 
fluctuations of a matter field in a thermal state \cite{BF}. In curved 
space the nontrivial geometry can give rise to interesting effects.
Much work has been done in the homogeneous and isotropic geometry of
cosmology and in black hole geometries \cite{CH95,TW,CCV,MV99,HS98,BFP00}.
Moving beyond the stress tensor, it will be seen that the same
considerations apply to correlators of any field. Recent examples of 
correlators in a locally de Sitter background include the vacuum 
polarization of scalar QED \cite{PTW}, the fermion self-energy of Yukawa 
theory \cite{PW}, and the self-mass-squared of a self-interacting scalar 
\cite{BOW}. The analysis of this paper allows one to be clear about the 
singular parts so that calculation of the finite parts may proceed 
unambiguously.

\centerline{\bf Acknowledgments}

This work was partially supported by NSF grants PHY-0244898 and PHY-0244714, 
and by the Institute for Fundamental Theory.


\begin{thebibliography}{99}

\bibitem{Barton}G. Barton, J. Phys. A {\bf 24}, 991 (1991);
{\bf 24}, 5563 (1991).

\bibitem{Eberlein} C. Eberlein, J. Phys. {\bf A 25}, 3015 (1992);
{\bf A 25}, 3039 (1992).

\bibitem{JR} M.T. Jaekel and S. Reynaud, Quantum Opt. {\bf 4}, 39 
(1992);
J. Phys. I France {\bf 2}, 149 (1992); {\bf 3}, 1 (1993); {\bf 3}, 
339 (1993).

\bibitem{WKF01}  C.-H. Wu, C.-I Kuo and L.H. Ford, Phys. Rev. A {\bf 65},
062102 (2002), quant-ph/0112056.

\bibitem{WF01} C.-H. Wu and L. H. Ford, Phys. Rev. D {\bf 64,} 045010 (2001),
quant-ph/0012144.

\bibitem{F82}L.H. Ford, Ann. Phys (NY) \textbf{144}, 238 (1982).

\bibitem{Kuo} C.-I Kuo and L.H. Ford, Phys. Rev. D \textbf{47}, 4510 
(1993), gr-qc/9304008.

\bibitem{BF} J. Borgman and L. H. Ford, Phys. Rev. D {\bf 70}, 064032 (2004),
gr-qc/\-0307043.

\bibitem{CH95} E. Calzetta  and B.L. Hu, Phys. Rev. D {\bf 49}, 6636 (1993),
gr-qc/9312036; Phys. Rev. D {\bf 52}, 6770 (1995), gr-qc/9505046.

\bibitem{TW} N. C. Tsamis and R. P. Woodard, Phys. Rev. {\bf D54}, 2621
(1996), hep-ph/9602317.

\bibitem{CCV}  E. Calzetta, A. Campos, and E. Verdaguer, Phys. Rev. D {\bf 56},
2163 (1997),  gr-qc/9704010.

\bibitem{MV99} R. Martin and E. Verdaguer, Phys. Rev. D {\bf 60}, 084008 
(1999), gr-qc/9904021.

\bibitem{HS98}  B. L. Hu and  K. Shiokawa, Phys. Rev. D {\bf 57}, 3474 
(1998), gr-qc/9708023.

\bibitem{BFP00}  C. Barrabes, V. Frolov, and R. Parentani, Phys. Rev. D 
{\bf 62}, 044020 (2000), gr-qc/0001102.

\bibitem{FW03} L. H. Ford and C.-H. Wu, Int. J. Theor. Phys. {\bf 42}, 15
(2003), gr-qc/0102063.

\bibitem{FJL} D.Z. Freedman, K. Johnson and J.I. Latorre, Nucl. Phys. 
{\bf B371}, 353 (1992).

\bibitem{JS} J. Schwinger, J. Math. Phys. {\bf 2}, 407 (1961).

\bibitem{M} K. T. Mahanthappa, Phys. Rev. {\bf 126}, 329 (1962).

\bibitem{BM} P. M. Bakshi and K. T. Mahanthappa, J. Math. Phys. {\bf 4}, 1
(1963); J. Math. Phys. {\bf 4}, 12 (1963).

\bibitem{K} L. V. Keldysh, Sov. Phys. JETP {\bf 20}, 1018 (1965).

\bibitem{CSHY} K. C. Chou, Z. B. Su, B. L. Hao and L. Yu, Phys. Rept. 
{\bf 118}, 1 (1985).

\bibitem{RJ} R. D. Jordan, Phys. Rev. {\bf D33}, 444 (1986).

\bibitem{CH} E. Calzetta and B. L. Hu, Phys. Rev. {\bf D35}, 495 (1987).

\bibitem{HV} G. 't Hooft and M. Veltman, Ann. Inst. Henri Poincar\'e,
{\bf XX}, 69 (1974).

\bibitem{OW} V. K. Onemli and R. P. Woodard, Class. Quant. Grav. {\bf 19},
4607 (2002), gr-qc/0204065.

\bibitem{PTW} T. Prokopec, O. T\"ornkvist and R. P. Woodard, Ann. Phys.
{\bf 303}, 251 (2003), gr-qc/0205130.

\bibitem{PW} T. Prokopec and R. P. Woodard, JHEP {\bf 0310}, 059 (2003),
astro-ph/0309593.

\bibitem{CV} A. Campos and E. Verdaguer, Phys. Rev. {\bf 49}, 1816 (1994),
gr-qc/9307027.

\bibitem{MV} R. Martin and E. Verdaguer, Phys. Rev. {\bf D61}, 124024 
(2000), gr-qc/0001098.

\bibitem{BOW} T. Brunier, V. K. Onemli and R. P. Woodard, Class. Quant.
Grav. {\bf 22}, 59 (2005), gr-qc/0408080.

\end{thebibliography}
\end{document}